\documentclass{elsart}
\usepackage{amsmath}
\DeclareMathAlphabet\cal{U}{eus}{m}{n}
\usepackage{graphicx}

\newcommand{\prev}{Phys. Rev.}
\newcommand{\prl}{Phys. Rev. Lett.}
\newcommand{\plett}{Phys. Lett.}
\newcommand{\nucph}{Nucl. Phys.}
\newcommand{\zeit}{Z. Phys.}

\newcommand{\ie}{i.e.}
\newcommand{\eg}{e.g.}
\newcommand{\cpi}{\mathrm{i}}
\newcommand{\expe}{\mathrm{e}}
\newcommand{\curl}{\mathrm{curl}}

\begin{document}

\begin{frontmatter}

\title{Generalized routhian calculations within \\ the Skyrme--Hartree--Fock 
approximation}

\author{D.~Sams\oe{}n},
\author{P.~Quentin}
\address{Centre d'Etudes Nucléaires de Bordeaux-Gradignan (IN2P3-CNRS 
and Univ. Bordeaux-I), Gradignan, France}
\author{J.~Bartel}
\address{Institut de Recherches Subatomiques (IN2P3-CNRS and Univ. 
Louis Pasteur), Strasbourg, France}

\begin{abstract}
We consider here variational solutions in the Hartree--Fock approximation upon 
breaking time reversal and axial symmetries.
When decomposed on axial harmonic oscillator functions, the corresponding 
single particle triaxial eigenstates as functions of the usual cylindrical 
coordinates ($r$, $\theta$, $z$) are evaluated on a mesh in $r$ and $z$ to be 
integrated within Gauss--Hermite and Gauss--Laguerre approaches and as Fourier 
decompositions in the angular variable $\theta$.
Using an effective interaction of the Skyrme type, the Hartree--Fock 
hamiltonian is also obtained as a Fourier series allowing a two dimensional 
calculation of its matrix elements.
This particular choice is shown to lead in most cases to shorter computation 
times compared to the usual decomposition on triaxial harmonic oscillator 
states.
We apply this method to the case of the semi-quantal approach of large 
amplitude collective motion corresponding to a generalized routhian formalism 
and present results in the $A=150$ superdeformed region for the coupling of 
global rotation and intrinsic vortical modes in what is known after 
Chandrasekhar as the \emph{S}-ellipsoid coupling case.
\end{abstract}

\end{frontmatter}

\section{Introduction}
\label{sec:intro}
In microscopic calculations within the Hartree--Fock approximation, one-body
eigenstates wave functions can be either evaluated numerically on a mesh or
decomposed on some truncated set of orthogonal functions. In the former case,
it is necessary to solve the static Schr\"o{}dinger-like integro-differential
or in the Skyrme case partial-derivative equations which can prove to be a
long process, whereas in the latter, one has to compute matrix elements of the
Hartree--Fock hamiltonian on the set and then diagonalize the obtained matrix.
This second method is known to be rather efficient in terms of computation
time. However, as one is bound to use only a truncated basis, the obtained
results are dependent on the basis parameters and size. Nevertheless the
energy difference between the truncated solution and the exact one (which
would correspond to an infinite basis and can be approached by the former
method) may in some cases be evaluated (see \eg{} \cite{Troncature}).

In order to describe a compact system of particles interacting through mostly
attractive two body forces, one often chooses as a basis set, eigenfunctions
of a deformed harmonic oscillator single particle hamiltonian. One truncates
it for instance \cite{Troncature} by only retaining the eigenstates whose
eigenenergies are lower than some truncation energy reference. The obtained
solutions thus depend on both the truncation energy and the deformation
parameters of the harmonic oscillator basis. For a given truncation, the
deformation parameters to be used are obtained by a minimization of the energy
with respect to these parameters.
When describing axially symmetric nuclei, it is appropriate to use axially
symmetric harmonic oscillator wave functions, all the calculations being then
performed in two dimensions.  To the best of our knowledge, when projecting
triaxially shaped solutions on a single particle basis, only triaxial harmonic
oscillator wave functions have been used \cite{Egido,Girod,Dudek}.
We present here an alternative way to describe triaxial shapes by using an 
axially symmetrical basis. Axially symmetric wave functions are eigenstates of 
the projection of the angular momentum on the quantization axis. By mixing 
states with different eigenvalues for this operator, one can obtain triaxial 
wave functions. As a matter of fact, in the position representation, those 
functions are expressed for any point of the $(r,z)$ plane as a complex 
Fourier series in the angular variable $\theta$.

Using, as we will do here, an effective interaction of the Skyrme type (namely
in the SkM$^{*}$ parametrisation \cite{Skmstar}), the total energy of the
nucleus is obtained by integrating an energy density functional which, as well
as the potentials of the Hartree--Fock hamiltonian, is expressed in terms of
sums, products and integer and non-integer powers of some local
densities. With our choice of axially symmetric basis states as we will show,
those densities, and hence the Hartree--Fock potentials and energy functional,
are obtained as real Fourier series in the angular variable.
By integrating the energy functional over the whole space, only the first term
of the Fourier series (which remains constant as a function of $\theta$) is
non vanishing. This component is calculated for each point in the $(r,z)$
plane and then integrated in two dimensions.  Using selection rules, the
calculation of the Hartree--Fock hamiltonian matrix elements can also be
performed through a single two-dimensional integral whereas when using
triaxial wave functions, all these integrations would be three-dimensional. Our
alternative method thus is expected to lead to shorter computation times 
provided that the number of axially symmetric basis wavefunctions necessary
for an accurate description of the triaxial solution is not too large.

Even for triaxial shapes, high order Fourier components are exactly zero and
the number of non vanishing components is determined by the truncation
parameter. In practical cases, one can impose a cutoff in the maximal order of
the retained components, due to the observed fast convergence of the numerical
results with respect to this maximal order.
For axially symmetric nuclei, only the first component of the densities are
non vanishing. We are thus able to describe both axially symmetric and
triaxial shapes within the same code by simply adjusting the maximal order of
the Fourier decompositions.
For largely triaxial shapes we can expect relatively important truncations 
effects within our method, which will lead to a reduced competitivity of our 
approach with respect to those using triaxial basis.

This calculation method has been applied to physical situations of triaxial
shapes combined with time-reversal symmetry breaking as obtained within the
well known cranking (or routhian) approximation or within approaches
corresponding to a generalization of the latter. This generalization which
will be called below the generalized routhian formalism has been briefly
sketched in a previous paper \cite{Nupha} and will be presented in more
details in Ref.~\cite{Routhgene}. It allows us to describe the dynamics of a
class of collective modes defined by a velocity field in a semi-quantal
approach \`a la WKB. It reduces to the addition of a time-odd constraint
$-\boldsymbol\beta\cdot\boldsymbol p$ to the static hamiltonian, where $\boldsymbol\beta$ is the
classical collective velocity field under study and $\boldsymbol p$ is the single
particle momentum operator.
The possible choices for the $\boldsymbol\beta$-fields have been restricted by
imposing the routhian eigenstates to be also eigenstates of the parity and of
the signature operator with respect to the first axis. 

This paper will be organized as follows. In Section \ref{sec:hfsrouth} we will
present our formalism with some calculation and computation details. Section
\ref{sec:tests} will be devoted to discussions about truncation effects and
the computation time we expect to gain and will end with the comparison of our
results with the literature. We will then turn to the generalized routhian
formalism in the \emph{S}-type ellipsoid case and present some preliminary
results in section \ref{sec:styp} before concluding and presenting
perspectives of future work in section \ref{sec:conclu}.

\section{Hartree--Fock--Skyrme generalized routhian}
\label{sec:hfsrouth}

\subsection{One and N-body routhians}
\label{ssec:routh}

Within classical mechanics, the canonical form of the equations of motion is
conserved under a canonical transformation of the dynamical variables by
adding to the hamiltonian the time derivative of a generating function. In the
case of a local transformation (\ie{} involving only the coordinates), this
time derivative writes as
\begin{equation}
\frac{\partial G}{\partial t} = - \boldsymbol\beta \cdot \boldsymbol p ,
\label{generating}
\end{equation}
where $\boldsymbol\beta$ is the collective velocity related to the transformation.
In the quantum mechanical case, it can be shown \cite{Routhgene} using a
certain class of unitary transformations that the density matrix solution of
the Hartree--Fock equation involving the generalized routhian
\begin{equation}
R = H - \frac{1}{2} ( \boldsymbol\beta \cdot \boldsymbol p + \boldsymbol p \cdot \boldsymbol\beta ) \equiv 
H - \boldsymbol\beta \cdot \boldsymbol p + \frac{\cpi\hbar}{2} \mathrm{div} \boldsymbol\beta .
\label{routhian}
\end{equation}
in lieu of the hamiltonian $H$ exhibits in the laboratory frame a dynamical
content which corresponds to the collective velocity field $\boldsymbol\beta$. When
spin degrees of freedom are present, it is necessary to also consider the
spin-rotation collective mode. Considering a vortical velocity field, the
routhian then writes
\begin{equation}
R = H - \boldsymbol\beta\cdot\boldsymbol p - \frac{\hbar}{2} \boldsymbol\Omega \cdot \boldsymbol\sigma ,
\label{routh-spin}
\end{equation}
where $\boldsymbol\sigma$ is the vector whose components are the three Pauli
matrices. As an example, in the case of global rotation where the collective
field is 
\begin{equation}
\boldsymbol\beta_{\mathrm{rot}} = \boldsymbol\Omega \times \boldsymbol r ,
\label{rot-field}
\end{equation}
the routhian is expressed by
\begin{equation}
R = H - \boldsymbol\beta_{\mathrm{rot}} \cdot\boldsymbol p - \frac{\hbar}{2} \boldsymbol\Omega 
\cdot \boldsymbol\sigma \equiv H - \boldsymbol\Omega \cdot (\boldsymbol\ell + \boldsymbol s) ,
\label{routh-rot}
\end{equation}
which is the well known cranking approximation.

Using Skyrme interaction, the routhian expectation value for a Slater 
determinant can be written as the integral over the whole space
\begin{equation}
\langle R \rangle = \int \left( {\cal H} (\boldsymbol r) - \hbar \boldsymbol\beta\cdot\boldsymbol j 
- \frac{\hbar}{2} \boldsymbol\Omega \cdot \boldsymbol\rho \right) \mathrm{d}\boldsymbol r
\label{routh-functional}
\end{equation}
where ${\cal H}$ is the energy density functional, $\boldsymbol j$ is the
current density and $\boldsymbol\rho$ is the spin-vector density whose
expressions are given in Appendix \ref{app:densities}.

Minimizing this expectation value with respect to all single particle wave
functions, one obtains the following expression for the one-body routhian
(see Ref.~\cite{Engel} for instance)
\begin{eqnarray}
h_q & = & U_q - \frac{\hbar^2}{2m} \boldsymbol\nabla f_q \cdot 
\boldsymbol\nabla + \frac{\cpi\hbar}{2} \left( \boldsymbol\alpha_q \cdot \boldsymbol\nabla + 
\boldsymbol\nabla \cdot \boldsymbol\alpha_q \right) \nonumber \\
\label{routh-1body}
&& - \hbar \left( \boldsymbol S_q + \cpi\hbar\boldsymbol\nabla V^{\mathrm{so}}_q \times 
\boldsymbol\nabla \right) \cdot \boldsymbol\sigma ,
\end{eqnarray}
where  $q$ stands for the considered charge state. The expressions of the 
various form factors entering this routhian are given in Appendix 
\ref{app:fields}.

\subsection{Symmetries}
\label{ssec:sym}

As said in the introduction, we have chosen here to use axially symmetric
harmonic oscillator wave functions for our basis states. Such wave functions
have been widely used \cite{Vautherin} to describe axially symmetric
variational solutions. For the single particle hamiltonian, they entail
symmetries with respect to the parity and the third component of the angular
momentum, and can be written in the coordinates representation in terms of
normalized Hermite and associated Laguerre polynomials $H_{n_z}$ and
$L_{n_r}^{|\Lambda|}$ (see Ref.~\cite{Abramowitz}) as
\begin{equation}
\varphi_\mu(\boldsymbol r) = \left[ \frac{\beta_z\beta_\perp^2}{\pi} 
\textrm{e}^{-(\xi^2+\eta)} \right]^{1/2} \expe^{\cpi\Lambda\theta} 
\eta^{|\Lambda|/2} H_{n_z}(\xi) L_{n_r}^{|\Lambda|}(\eta) ,
\label{basis-func}
\end{equation}
where $\mu$ represents the set $(n_z n_r \Lambda)$ of quantum numbers,
$\beta_z$ and $\beta_\perp$ are the usual oscillator constants
\cite{Vautherin} which are generally related to the parameters
$\beta_0=(\beta_z\beta_\perp^2)^{1/3}$ and $q = (\beta_\perp/\beta_z)^2$ and
$\xi$ and $\eta$ are given in terms of the cylindrical coordinates $r$ and $z$
by
\begin{equation}
\xi = z \beta_z , \qquad\qquad \eta = r^2 \beta_\perp^2 .
\label{reduced-coord}
\end{equation}
For notational convenience, we will omit in the following the $\xi$ and $\eta$ 
dependence of the polynomials.

Whenever the time-reversal symmetry is broken, the third component of the
angular momentum can no longer be chosen as a symmetry. It is well known
however, that the Hartree--Fock hamiltonian obtained from the Skyrme
interaction, the parity and for instance the first component of the signature
operator \cite{Goodman} defined as
\begin{equation}
S_1 = \Pi_2 \Pi_3 \sigma_1 ,
\label{signature}
\end{equation}
(where $\Pi_i$ is the reflection operator in the $i$ direction, and $\sigma_1$
is the usual Pauli matrix) form a set of commuting operators. Imposing that
the routhian also commutes with the two last symmetry operators limits the
possible choice for $\boldsymbol\beta$. As an example, if we restrict ourselves to
first order polynomials in the coordinates, the vortical velocity field given
by its components
\begin{equation}
\begin{array}{rcl}
\beta_1 & = & 0 , \\
\beta_2 & = & a x_3 , \\
\beta_3 & = & b x_2 
\end{array}
\label{possible-alpha}
\end{equation}
is the most general one that fulfills these commutation requirements.
This analytical form corresponds to the so-called Riemann \emph{S}-type 
ellipsoid solution presented by Chandrasekhar \cite{Chandra} in the context of 
fluid dynamics, whose application to nuclear physics has been discussed in 
particular in \cite{Nupha,Staggering}. This field can then be rewritten as the 
sum of a global rotation term of angular velocity $\Omega$ and an intrinsic 
vorticity term of angular velocity $\omega$ both perpendicular to the first 
axis in the form
\begin{equation}
\begin{array}{rcl}
\beta_1 & = & 0 , \\
\beta_2 & = & -(\Omega + \omega/q) x_3 , \\
\beta_3 & = & (\Omega + \omega q) x_2 , 
\end{array}
\label{stype-field}
\end{equation}
where $q$ is the $a_3/a_2$ axis ratio of the ellipsoid approximating the
nuclear shape. The coupling of the rotation perpendicular to the first axis
with the spins is taken into account by a term proportional to
$\Omega\sigma_1$ in the routhian which still commutes with the signature
operator.

Since the routhian single-particle eigenstates must have the same symmetries
as the routhian, it is appropriate that the basis states are also eigenstates
of the parity and signature operator. The action of the parity $P$ and
signature $S_1$ operator on axial harmonic oscillator states is given in usual
notations by
\begin{eqnarray}
\label{p-action}
P |n_z n_r \Lambda \Sigma\rangle &=& (-)^{n_z+\Lambda} | n_z n_r \Lambda
\Sigma \rangle , \\
\label{s1-action}
S_1 |n_z n_r \Lambda \Sigma\rangle &=& (-)^{n_z} | n_z n_r -\Lambda -\Sigma  
\rangle ,
\end{eqnarray}
from where we can see that $S_1{}^2$ is the identity operator, and thus the
eigenvalues of $S_1$ are $\pm 1$. We then obtain eigenstates of the parity and
signature in the form
\begin{equation}
|\mu s \rangle = \frac{\sqrt{2}}{2} \left[ |\mu \half \rangle + s 
(-)^{n_z} |\bar\mu -{}\half \rangle \right] ,
\qquad s = \pm 1 ,
\label{eigenstates}
\end{equation}
where $\mu$ ($\bar\mu$) represent the set $(n_z, n_r, {}\Lambda)$ [$(n_z, n_r,
-{}\Lambda)$] of quantum numbers, and we have
\begin{equation}
P |\mu s\rangle = (-)^{n_z+\Lambda} |\mu s\rangle , \qquad
S_1 |\mu s\rangle = s |\mu s\rangle .
\label{eigenvalues}
\end{equation}

The single-particle eigenstates obtained after minimizing the routhian are 
written as
\begin{equation}
|k\rangle = \sum_\mu C_\mu^k |\mu s\rangle ,
\label{routh-eigen}
\end{equation}
where the sum runs over basis states having the same eigenvalue for the two
symmetry operators. In the coordinates representation, these eigenstates can
be decomposed in two components with spin $\half$ and $-{}\half$ as
\begin{equation}
\Phi_k(\boldsymbol r) = \Psi_k^+(\boldsymbol r) |\half\rangle + s \Psi_k^-(\boldsymbol r) 
|-{}\half\rangle ,
\label{eigenstate}
\end{equation}
and one deduces from equations (\ref{basis-func}) and (\ref{routh-eigen}) 
\begin{equation}
\Psi_k^\pm (\boldsymbol r) = \left[ \frac{\beta_z\beta_\perp^2}{2\pi} 
\textrm{e}^{-(\xi^2+\eta)} \right]^{1/2} \sum_\mu(\pm)^{n_z} C_\mu^k 
\expe^{\pm i\Lambda\theta} \eta^{|\Lambda|/2} H_{n_z} L_{n_r}^{|\Lambda|} .
\label{spineur}
\end{equation}
We are now able to calculate the expressions of the various densities of 
Appendix \ref{app:densities}. The density $\rho$ of equation(\ref{ro-dens}) 
for instance rewrites with the notations of (\ref{eigenstate}) as
\begin{equation}
\rho = \sum_k {\Psi_k^+}^*\Psi_k^+ + {\Psi_k^-}^*\Psi_k^-
\label{ro-dens-2}
\end{equation}
since $s^2=1$. Now inserting the expression (\ref{spineur}) and rearranging
the terms we get
\begin{eqnarray}
\rho & = & \frac{\beta_z\beta_\perp^2}{2\pi} \textrm{e}^{-(\xi^2+\eta)}
\sum_{\mu,\mu'} \left(\sum_k {C_\mu^k}^*C_{\mu'}^k \right) 
\eta^{(|\Lambda|+|\Lambda'|)/2} H_{n_z} L_{n_r}^{|\Lambda|} H_{n'_z}
L_{n'_r}^{|\Lambda'|} \nonumber \\
\label{ro-dens-3}
&& \left( \expe^{i(\Lambda'-\Lambda)\theta} + 
(-)^{n_z+n_z'}\expe^{i(\Lambda-\Lambda')\theta} \right) .
\end{eqnarray}
The sum over $k$ only involves the components $C_\mu^k$ of (\ref{routh-eigen}) 
and is nothing but the matrix element of the density matrix, that is
\begin{equation}
\sum_k {C_\mu^k}^* C_{\mu'}^k = \rho_{\mu\mu'} .
\label{ro-matrix}
\end{equation}
In the general case this matrix is hermitian, but it can be shown
\cite{MaThese} that the choice made for the basis states makes it a real
symmetrical matrix. Using this property, the imaginary parts of the $\mu\mu'$
and $\mu'\mu$ terms in equation (\ref{ro-dens-3}) cancel each other, and we
can write
\begin{eqnarray}
\rho & = & \frac{\beta_z\beta_\perp^2}{2\pi} \textrm{e}^{-(\xi^2+\eta)}
\sum_{\mu,\mu'} \rho_{\mu\mu'}
\eta^{(|\Lambda|+|\Lambda'|)/2} H_{n_z} L_{n_r}^{|\Lambda|} H_{n'_z}
L_{n'_r}^{|\Lambda'|} \nonumber \\
\label{ro-dens-4}
&& \cos [(\Lambda-\Lambda')\theta] \left( 1 + (-)^{n_z+n_z'} \right) .
\end{eqnarray}
Finally, since the basis states $|\mu\rangle$ and $|\mu'\rangle$ have the same 
eigenvalue for the parity operator we can deduce from (\ref{eigenvalues}) that
\begin{equation}
(-)^{n_z+n'_z} = (-)^{\Lambda-\Lambda'} ,
\label{same-eigen}
\end{equation}
and the $\mu\mu'$ term of the sum (\ref{ro-dens-4}) vanishes if the difference 
$\Lambda-\Lambda'$ is an odd integer. We then write at last the density 
$\rho$ as
\begin{eqnarray}
\rho & = & \frac{\beta_z\beta_\perp^2}{\pi} \textrm{e}^{-(\xi^2+\eta)}
\sum_{\mu,\mu'} \rho_{\mu\mu'}
\eta^{(|\Lambda|+|\Lambda'|)/2} \delta^{2p}_{\Lambda-\Lambda'} 
\cos [(\Lambda-\Lambda')\theta] \nonumber \\
\label{ro-dens-last}
&& H_{n_z} H_{n'_z} L_{n_r}^{|\Lambda|} L_{n'_r}^{|\Lambda'|} . 
\end{eqnarray}
with the definition
\begin{equation}
\delta^{2p}_{\Lambda-\Lambda'} = \begin{cases}
1 &\quad \text{if }\Lambda-\Lambda' 
\text{ is an even integer} , \\
0 &\quad \text{if }\Lambda-\Lambda' \text{ is an odd integer} .
\end{cases}
\label{cronecker-1}
\end{equation}

The calculations of the other local densities entering the one-body routhian
are presented in Appendix \ref{app:denscalc}. It is clear from equation
(\ref{ro-dens-last}) and from the results of this Appendix that all the
densities are obtained as real Fourier series in the angular variable
$\theta$.  As some of the Fourier components of these densities are vanishing,
we recall in Table~\ref{dens-decomp} the non-vanishing components together
with their parity with respect to the $z$ variable. The maximal order of the
non-vanishing components is $2\Lambda_\mathrm{max}+1$ where
$\Lambda_\mathrm{max}$ is the higher $\Lambda$ value among the basis
states which is equal to $N_0$ in the case of a spherical basis and increases
for deformed basis.

\begin{table}[t]
\begin{center}
\caption{Definition of the non-vanishing components of the Fourier series
representing the various local densities in use; their parity $\pi_z$ with
respect to $z$ is also given.}
\bigskip
\label{dens-decomp}
\begin{tabular}{c|c c}
Densities & function & $\pi_z$ \\ \hline
$\rho$, $\tau$, $\nabla^2\rho$, $\boldsymbol\nabla\cdot\boldsymbol J$ & $\cos 2p\theta$ & +\\
$j_r$, $\curl_r\boldsymbol\rho$ & $\sin [(2p+1)\theta]$ & - \\
$j_\theta$, $\curl_\theta\boldsymbol\rho$, $\curl_z\boldsymbol j$, $\rho_z$ & $\cos
[(2p+1)\theta]$ & - \\
$j_z$, $\curl_z\boldsymbol\rho$, $\curl_\theta\boldsymbol j$, $\rho_\theta$ & $\sin
[(2p+1)\theta]$ & + \\
$\curl_r\boldsymbol j$, $\rho_r$ & $\cos [(2p+1)\theta]$ & + \\ \hline
\end{tabular}
\bigskip
\end{center}
\end{table}

The form factors of the one-body routhian are expressed in 
Appendix~\ref{app:fields} as sums and products of these densities, non-integer powers
of $\rho$ and the Coulomb potential. We show in Appendix~\ref{app:coul} how
the Coulomb potential can be obtained as a Fourier series in the $\theta$
variable, and the Fourier decompositions of the non-integer powers of $\rho$
are obtained through weighted integrals over $\theta$ computed with a 48 points
Gauss--Legendre quadrature formula. The Fourier decomposition of $\boldsymbol\beta$ 
is readily obtained from equation (\ref{stype-field}) as
\begin{equation}
\begin{array}{rcl}
\beta_r & = & z(\Omega + \omega q) \sin\theta , \\
\beta_\theta & = & -z(\Omega + \omega q) \cos\theta , \\
\beta_z & = & -r(\Omega + \omega/q) \sin\theta .
\end{array}
\end{equation}
Using the well known relations expressing products of trigonometrical
functions as sums of trigonometrical functions of the sum and difference of
their arguments, it is then straightforward to obtain the routhian form
factors as Fourier series, whose non-vanishing components are given in
Table~\ref{pot-decomp}.

\begin{table}[t]
\begin{center}
\caption{Same as Table \ref{dens-decomp}, but for the one-body routhian form
factors.}
\bigskip
\label{pot-decomp}
\begin{tabular}{c|c c}
Form factors 		  & function 		& $\pi_z$ \\ \hline
$U$, $f$, $V^\mathrm{so}$ & $\cos 2p\theta$	& + \\
$\alpha_r$ 		  & $\sin (2p+1)\theta$ & - \\
$\alpha_\theta$, $S_z$ 	  & $\cos (2p+1)\theta$ & - \\
$\alpha_z$, $S_\theta$ 	  & $\sin (2p+1)\theta$ & + \\
$S_r$ 			  & $\cos (2p+1)\theta$ & + \\ \hline
\end{tabular}
\bigskip
\end{center}
\end{table}

Having expressed the routhian form factors as Fourier series in the angular
variable allows us to calculate the routhian matrix elements as
two-dimensional integrals, using simple selection rules. Indeed, if we write
for instance, referring to Table~\ref{pot-decomp}, the scalar potential $U$ as
\begin{equation}
U(\eta,\theta,\xi) = \sum_{n\geq 0} \cos (2n\theta) U^{(2n)}(\eta,\xi) ,
\label{decomp-U}
\end{equation}
the part of the matrix element involving this form factor is spin diagonal and
writes
\begin{equation}
\langle \mu s | U | \mu' s \rangle = \frac{1}{2} \sum_{n\geq0} \langle \mu
| \cos (2n\theta)\, U^{(2n)} | \mu' \rangle
+ (-)^{n_z+n'_z} \langle \bar\mu | \cos (2n\theta)\, U^{(2n)}
| \bar\mu' \rangle ,
\label{mat-U-1}
\end{equation}
readily performing the spin part of the scalar product.
For a given $n$, the angular part of the first term on the right-hand side is 
simply proportional to 
\begin{equation}
\int_0^{2\pi} \expe^{\cpi(\Lambda'-\Lambda)\theta} \cos m\theta = \begin{cases}
0 \phantom{\pi(1+\delta_{|\Lambda-\Lambda'|}^0)} 
\quad &\text{if } m \neq |\Lambda-\Lambda'| , \\
\pi (1+\delta_{|\Lambda-\Lambda'|}^0)\phantom{0} \quad &\text{if }m =
|\Lambda - \Lambda'| ,
\end{cases}
\label{select-rule}
\end{equation}
and is identical to the angular part of the second term up to a complex 
conjugation. It can then be easily seen that equation (\ref{mat-U-1}) reduces 
to 
\begin{equation}
\langle \mu s | U | \mu' s \rangle = \langle \mu \half | \cos
(|\Lambda-\Lambda'|\theta)\, U^{(|\Lambda-\Lambda'|)} | \mu' \half \rangle ,
\label{mat-U-2}
\end{equation}
which writes in analytical form
\begin{eqnarray}
\langle\mu s | U | \mu' s \rangle &=& 
(1 + \delta_{\Lambda-\Lambda'}^0) \int_0^\infty \mathrm{d}\eta \int_0^{\infty}
\mathrm{d}\xi \expe^{-(\xi^2+\eta)} \eta^{(|\Lambda|+|\Lambda'|)/2} 
\nonumber \\
\label{mat-U-last}
&& U^{(|\Lambda-\Lambda'|)} H_{n_z} H_{n'_z} L_{n_r}^{|\Lambda|} L_{n'r}
^{|\Lambda'|} , 
\end{eqnarray}
where we have restricted the $\xi$ integration to positive values, since the
integrand is an even function of $\xi$. The calculation of the other terms of
the routhian matrix elements also reduces to two-dimensional integrals and is
presented in Appendix \ref{app:matrix}. These integrations are performed
through 10 points Gauss--Hermite and Gauss--Laguerre integrations (for the
Gauss--Hermite integration, we use in fact a 20 points method which reduces to
10 points due to $\Pi_3$ symmetry properties). We therefore only need to
compute the values of these densities and routhian form factors on the mesh
points of a 10 by 10 grid.

\subsection{Calculation of some observables}
\label{ssec:calcul}

Within the Skyrme--Hartree--Fock formalism, the calculation of observables is
generally reduced to a three-dimensional integral as is obviously the case for
the routhian expectation value in equation
(\ref{routh-functional}). Expressing this integral as the integral of a
Fourier series, only the first component needs to be integrated --- which is
done through a two-dimensional integral --- since the integrals of the others
are vanishing. For instance, the root mean square radius of the mass
distribution which is given by
\begin{equation}
R^2_{\mathrm{rms}} = \frac{1}{A} \int {\boldsymbol r}^2 \rho(\boldsymbol r) \;\mathrm{d}\boldsymbol
r
\label{obs-rms}
\end{equation}
where $A$ is the total number of nucleons, reduces to the two-dimensional 
integral
\begin{equation}
R^2_{\mathrm{rms}} = \frac{4\pi}{A}\int_0^\infty r \;\mathrm{d}r
\int_0^\infty\mathrm{d}z \;(r^2 + z^2) \rho^{(0)}(r,z) ,
\label{obs-rms-2}
\end{equation}
since $\boldsymbol r^2$ does not depend upon $\theta$. The quadrupole operator
expectation value is easily obtained as
\begin{equation}
\langle Q_{0} \rangle = 4\pi\int_0^\infty r \;\mathrm{d}r
\int_0^\infty \;\mathrm{d}z (2z^2 - r^2) \rho^{(0)}(r,z) ,
\label{obs-q20}
\end{equation}
similarly to what has been done for the root mean square radius calculation.
The $a_2$ semi-axis length is given up to some normalization factors as the 
square root of $\langle {x_2}^2 \rangle$ written as
\begin{equation}
\langle {x_2}^2 \rangle = \frac{1}{A} \int r^2\sin^2\theta \rho(\boldsymbol r)\;
\mathrm{d}\boldsymbol r .
\label{obs-axis-1}
\end{equation}
The integrand easily writes as a Fourier series upon developing $\sin^2
\theta$, and using the results of Table~\ref{dens-decomp}. Retaining only its 
first component, we get
\begin{equation}
\langle {x_2}^2 \rangle = \frac{2\pi}{A} \int_0^\infty r \;\mathrm{d}r
\int_0^\infty\mathrm{d}z\; r^2 \left( \rho^{(0)}(r,z) - \half \rho^{(2)}(r,z)
\right) .
\label{obs-axis-2}
\end{equation}
And the expectation value of the non-axial quadrupole moment $Q_{22}$ is 
obviously
\begin{equation}
\langle Q_{22} \rangle = A \langle {x_2}^2 - {x_1}^2 \rangle = -2\pi
\int_0^\infty r \;\mathrm{d}r \int_0^\infty\mathrm{d}z \;r^2 \rho^{(2)}(r,z) .
\label{obs-q22}
\end{equation}

The first component of the total angular momentum is obtained from the current
and spin-vector densities as 
\begin{equation}
\langle{L_1}/{\hbar}\rangle = \langle (\boldsymbol r \times \boldsymbol j + \boldsymbol\rho)
\cdot \boldsymbol e_1 \rangle ,
\label{obs-angular-1}
\end{equation} 
where $\boldsymbol e_1$ is the unit vector in the first direction. It is expanded as 
\begin{equation}
\langle L_1/\hbar\rangle = \int \left[ rj_z\sin\theta -z(j_\theta\cos\theta +
j_r\sin\theta) + \half (\rho_r\cos\theta - \rho_\theta\sin\theta) \right]
\mathrm{d}\boldsymbol r ,
\label{obs-angular-2}
\end{equation}
and finally, referring to Table~\ref{dens-decomp} we get,
\begin{equation}
\langle L_1/\hbar\rangle = \frac{2\pi}{A} \int_0^\infty r \;\mathrm{d}r
\int_0^\infty \mathrm{d}z \left[ rj_z^{(1)} -z(j_\theta^{(1)} + j_r^{(1)}) +
\half (\rho_r^{(1)} - \rho_\theta^{(1)}) \right] .
\label{obs-angular-last}
\end{equation}
Taking into account the total angular momentum quantization rule for an even
number of nucleons, we will only retain solutions that satisfy
\begin{equation}
\langle L_1/\hbar \rangle^2 = I(I+1) ,
\label{angular-quant}
\end{equation}
where $I$ must be an even integer. The dynamical moment of inertia is
calculated numerically as the first derivative of this angular momentum with
respect to the angular velocity, namely
\begin{equation}
\Im^{(2)} = \frac{\partial I}{\partial \Omega} ,
\label{dynamic-mom}
\end{equation}
which, as mentioned by Gall \etal{} \cite{Gall-190}, provides a better 
numerical stability than the second derivative of the energy with respect to
$\Omega$.

\section{Numerical tests}
\label{sec:tests}

\subsection{Convergence and computation time}
\label{ssec:time}

Describing as we do here the single-particle states as a decomposition on a
truncated harmonic oscillator basis, it is then crucial to study carefully the
convergence of the solution with respect to the truncation.  We are using the
usual deformed truncation scheme, namely considering only as basis states those
having an energy lower than what is obtained for a state with $N_0$ quanta of
the equivalent spherical oscillator \cite{Vautherin}. Discussions on this
problem in the static case can be found in the literature \cite{Troncature}. 
We will thus concentrate here on the rotational case.

\begin{table}[b]
\begin{center}
\bigskip
\caption{Values of some observables in the $I=20$~$\hbar$ solution of the
nuclei \nuc{80}{Sr} as a function of $N_0$ (see Subsection~\ref{ssec:calcul}
for calculation details). We used the parameters $\beta_0=0.534$ and
$q=1.2658$.}
\bigskip
\label{zirco}
\begin{tabular}{c|c c c c c c|}
 & 6 & 8 & 10 & 12 & 14\\ \hline
$\langle R\rangle$ (MeV) &
-686.69 & -688.07 & -688.75 & -690.28 & -690.61 \\
$\langle Q_0\rangle$ (b) & 
   6.04  &    6.10  &    6.12  &    6.13  &    6.10  \\
$\langle Q_{22}\rangle$ (b) & 
   0.32  &    0.44  &    0.47  &    0.41  &    0.43  \\ \hline
$\Omega$ (MeV) & 
   0.831 &    0.831 &    0.831 &    0.836 &    0.832 \\ 
$\Im^{(2)}$ ($\hbar^2$.MeV$^{-1}$) & 
  24.46  &   24.06  &   23.96  &   23.93  &   23.91  \\ \hline
\end{tabular}
\end{center}
\end{table}

We have studied as an example rotating solutions of the nuclei \nuc{80}Sr. In
Table~\ref{zirco} we present some properties of the $I=20$~$\hbar$ solution
for $N_0$ varying between 6 and 14 major shells. The deformation parameters
were obtained from an optimization of the static solution. One sees that the
routhian expectation value which is converged up to 10 eV (for a given basis
size) vary by less than 0.23 percent as one adds two more shells. We have
checked that the convergence of the various terms entering the routhian
(eqs.~\ref{func-kin}-\ref{func-coul}) also converge with such a good accuracy,
with the noticeable exception of the time-odd contribution which is roughly
700 times smaller than the total routhian and varies by 5 percent between 8
and 10 shells.
Now looking at the deformation properties, we see a rather nice convergence of
the quadrupole moment above $N_0=8$. The variations of the non-axial
quadrupole moment are of the same order in absolute values.

In the lower part of Table \ref{zirco} we can see that the angular velocity
varies by less than 0.5 percent as two shells are added. The convergence of
the $\Im^{(2)}$ moment of inertia is even better, since its variation rate is
constantly decreasing. This good convergence of the moment of inertia can be
easily interpreted. Indeed, the moment of inertia is a differential quantity,
and it is reasonable to expect small truncation effects for it in the near
vicinity of a given solution.

Let us now discuss the computation time required by our formalism. First of
all, we present in Table \ref{time-1} the time needed to perform an iteration
in the axial and the triaxial case, that is in the limiting case where only
the first Fourier components are retained and in the case where all the
non-vanishing components are used. In the triaxial case, we give the values
for the first iteration where some initialization parameters are computed (see
Appendix \ref{app:coul}) and for the following iterations which are
shorter. We also present for comparison the computation time used by the
static (and axial) formalism of Vautherin \cite{Vautherin}.
In static cases, the ratio between our formalism and Vautherin's is somehow
small for low $N_0$ values, but almost reaches 20 for $N_0=14$. This is due to
our choice of the parity and signature symmetries for the single-particle
states which only split each parity block into two, in contrast to Vautherin's
case. The time reversal symmetry breaking also doubles the size of the basis.
On the contrary, excluding the initializations performed during the first
iteration, the ratio between the triaxial and axial case in our formalism
always decreases and stays relatively small.

\begin{table}[t]
\begin{center}
\caption{Computation time in seconds (measured on a HP9000/780 workstation), of
one Hartree--Fock iteration within various formalisms as a function of
$N_0$. In the triaxial case, the first iteration is referred by $^{(1)}$ and
the following by $^{(2)}$.}
\bigskip
\label{time-1}
\begin{tabular}{l|c c c c c c|}
 &  6   &    8   &   10   &   12   &   14\\ \hline
Vautherin's formalism & 
   0.24 &   0.59 &   1.46 &   3.07 &   5.98 \\
``axial'' case(parity,signature)  &
   0.57 &   1.87 &   7.31 &  32.8  & 108    \\
triaxial case \ $^{(1)}$ &
   9.65 &  15.6  &  28.8  &  68.5  & 169    \\
triaxial case \ $^{(2)}$ &
   1.49 &  4.69  &  15.2  &  52.1  & 150    \\ \hline
\end{tabular}
\bigskip
\end{center}
\end{table}

An almost direct comparison of our code could be performed with the one built
by Dobaczewski and Dudek \cite{Dudek-comp} using a triaxial harmonic
oscillator basis for the \nuc{152}{Dy} nucleus with the SkM$^{*}$ effective
force.  The basis size is defined by M of the order of 300 in their
notation. This corresponds roughly to a deformation dependent truncation with
$N_0 =10$ in our notation. These authors quote a time of 9 seconds on a CRAY
C90 for the vectorised version. They provide a factor of two to three to
translate into the time used on e.g. an IBM RS/6000 station for a non
vectorised version. The corresponding time in our calculation of \nuc{150}{Gd}
with roughly the same basis size and after the first iteration, would be of
less than 20 seconds which is of the order or slightly better than the quoted
time.

Another way to estimate the advantages, or at least the competitiveness of our
approach consists in extrapolating from our code the computation time required
by triaxial basis calculations. For instance, in our case the routhian matrix
element calculation involves two-dimensional integrations, where it would
involve three-dimensional integrations in the fully triaxial case. Using 10
points Gauss-like integration method we can expect the matrix elements to be
computed 10 times faster within our formalism. The same factor is expected for
the densities calculation.
In our formalism, each of the $N_0$ non-vanishing Fourier components of the
Coulomb potential are calculated on a 10 by 10 mesh in the $(r,z)$ plane
performing a 10 by 10 integration in the $(r',z')$ plane and a 48 points
integration in the angular variable $\theta$. In the fully triaxial case, it
has to be computed for each of the points in the $(x,y,z)$ mesh, each time
performing a three-dimensional integral with 10 points. We thus expect our
formalism to be roughly a factor $N_0/2$ slower than the fully triaxial case
during the first iteration.
The routhian diagonalization is also time-consuming but it should take the
same amount of time within the two formalisms. Finally it seems that the
computation of the energy and the routhian form factors should be slower in
our formalism, but the time required for those calculations is negligible as
compared to the overall execution time of an iteration.

All these remarks allow us to calculate an ``expected'' computation time for
the fully triaxial formalism. We present this expectation in Table \ref{time-2}
together with the computation time of the discussed parts of our formalism.
Comparing the last line of Tables \ref{time-1} and \ref{time-2}, we see that
our formalism is expected to be 4 times faster for $N_0=14$, and this factor
is likely to be even larger for lower $N_0$ values, reaching a factor of 9 for
$N_0=6$.
These factors result of course only from a guess, and should be confirmed by
existing fully triaxial codes.

\begin{table}[t]
\begin{center}
\caption{Computation time in seconds of various parts of our formalism during
the first iteration (measured on a HP9000/780 workstation). The last line 
gives an expectation of the time required by a fully triaxial formalism using 
triaxial wave functions.}
\bigskip
\label{time-2}
\begin{tabular}{l|c c c c c c|}
 &  6  &  8  & 10  & 12  & 14  \\ \hline
Diagonalization & 0.08 & 0.48 & 3.19 & 22.4 & 83.2 \\
Coulomb potential & 8.17 & 10.9 & 13.6 & 16.4 & 19.1 \\
Matrix elements \\
\hspace{2.5ex} and densities & 1.05 & 3.63 & 10.6 & 25.8 & 56.4 \\ \hline
Fully triaxial & 13.7 & 40.1 & 113 & 289 & 661 \\ \hline
\end{tabular}
\bigskip
\end{center}
\end{table}

Of practical interest also, would be a comparison between the computation
times needed in our case and in other competing and totally different
approaches to describe heavy nuclei within the same range of deformations. For
the same \nuc{150}{Gd} nucleus at the same deformation (ground-state $\beta$
value, $\gamma=15^\circ$) and for the same basis size ($N_0=8$) and iteration
number (30), the Bruy\`{e}res le Ch\^{a}tel code (discarding the time needed
to treat pairing correlations) would take \cite{Libert} about 30 percent more
time than our code. Of course, the Gogny effective force which is used there
is completely different from our Skyrme force.
Another point of comparison could be the 3D calculations of solutions having
the same symmetries for slightly heavier deformed nuclei ($A\sim190$) using
the imaginary-time step technique where the computing time for a similar
degree of convergence would be slightly larger \cite{Jacques} than
ours. However it should be stressed that in their case the time used to treat
pairing correlations has not been discarded.

It then appears that the orders of magnitude of the computing times for all
considered codes, are identical with a slight advantage of our method in
comparable approaches.  It could be noted finally that the rather recent
character of the present version of our code should reasonably leave some
place for further optimization.

\subsection{Results of superdeformed \nuc{150}{Gd}}
\label{ssec:gado}

We have chosen to test the physical results obtained within our formalism by
comparison with the Hartree--Fock results of Bonche \etal{} \cite{Bonche-Gado}
for the yrast band of the nucleus \nuc{150}{Gd} as calculated with the
SkM$^{*}$ parameterization. Their formalism is based upon the computation of
the one-body eigenstates on a three-dimensional mesh by solving the
Schr\"o{}dinger-like Hartree--Fock equation and thus are free from truncation
effects but of course are contingent upon mesh size and related approximate
numerical derivatives. The calculations presented here have been performed
using 10 major shells. Though we should take into account at least the 21
lower order Fourier components of the densities, we have checked that the
results are unchanged if we only use 7 Fourier components. This is due to the
very little triaxiality of the solutions whose maximal triaxial axis ratio is
only 1.02 (corresponding to $\gamma<1^\circ$) and is reached at the end of the
band.

\begin{figure}[t]
\begin{center}
\includegraphics[width=.75\textwidth]{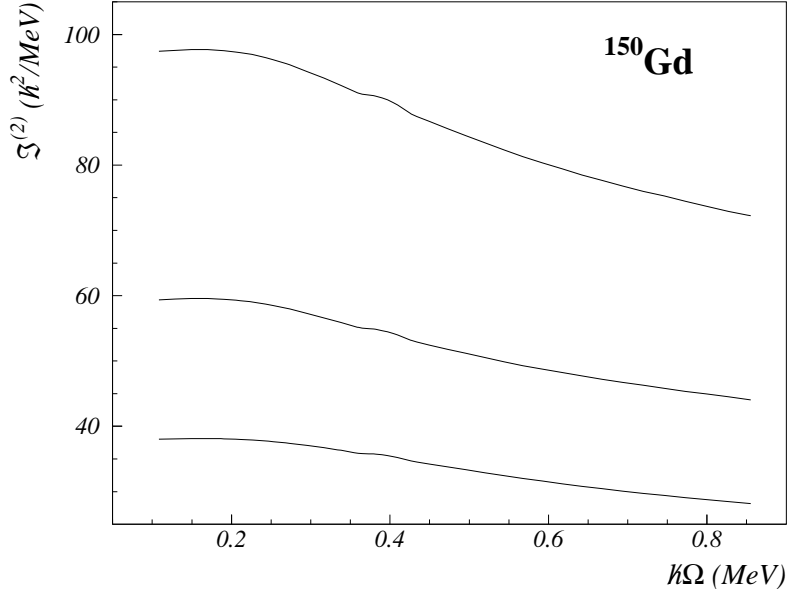}
\caption{Dynamical moment of inertia of the \nuc{150}{Gd} yrast band as a
function of the angular velocity. The two curves at the bottom of the figure
are the contribution of neutrons (upper curve) and protons (lower curve) to
the moment of inertia.}
\label{inertia}
\bigskip
\end{center}
\end{figure}

The basis parameters optimization has been performed in several steps. We have
obtained a first set of parameters from an optimization of a static solution
constrained to the experimental value of the charge quadrupole moment of the
yrast band. We have then minimized the routhian expectation values for
rotating solutions with angular momentum ranging from 10 to 80~$\hbar$ with
steps of 5~$\hbar$. Interpolating between these points, we obtained optimized
parameters with steps of 2~$\hbar$.

We finally minimized the routhian for each of these quantized values of the
angular momentum reaching an overall precision on the minimum of less than
0.1 keV. All these optimizations have been performed through successively
minimizing quadratic fits of the routhian in the $\beta_0$ and $q$ variables.
We only performed calculations for angular momentum above 10 $\hbar$ because
we did not want here to describe the transition from superdeformation to normal
deformation.

\begin{figure}[t]
\begin{center}
\includegraphics[width=.75\textwidth]{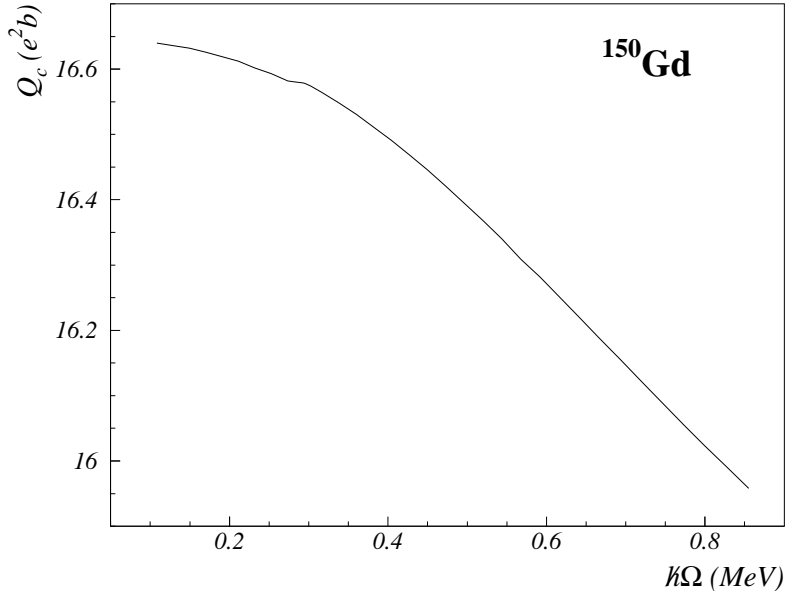}
\caption{Charge quadrupole moment of the \nuc{150}{Gd} yrast band as a function
of the angular velocity.}
\label{quadrupole}
\bigskip
\end{center}
\end{figure}

We show in Figs. \ref{inertia} and \ref{quadrupole} respectively the Dynamical
moment of inertia and the charge quadrupole moment (calculated by replacing the
matter density $\rho$ in eq.~(\ref{obs-q20}) with the proton matter density
$\rho_p$). While our values of the dynamical moment of inertia is in perfect
agreement with the one depicted in \cite{Bonche-Gado}, we obtain a charge
quadrupole moment which is slightly larger by 0.2 e$^2$b. This effect cannot
be related to truncation effects since we have checked that the same
quadrupole moments are obtained with $N_0=14$. However, the variation of this
moment over the yrast band is quantitatively well reproduced by our
calculations.

\begin{figure}[t]
\begin{center}
\includegraphics[width=.75\textwidth]{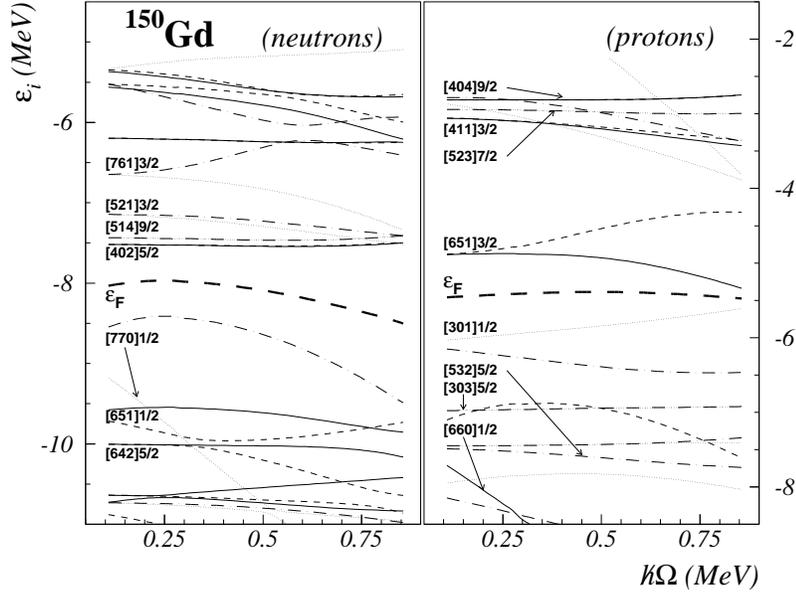}
\caption{Single-particle routhian spectra in the \nuc{150}{Gd} yrast band for
both protons and neutrons. The convention used for the (parity,signature)
representation is the following: (+,+) in full lines, (+,--) in dashed lines,
(--,+) in dash-dotted lines and(--,--) in dotted lines.}
\label{routhians}
\bigskip
\end{center}
\end{figure}

We finally plot in Fig. \ref{routhians} the neutron and proton single-particle
routhians. We are globally in good agreement with the results of
Ref.~\cite{Bonche-Gado} (see for instance the proton gap for $Z=66$ between
the [651] and [411] states and the intruder orbital in the proton spectra
appearing on top of the spectra above $\hbar\Omega= 0.5$~MeV) but we observe
some small differences. Among the less important are the labels of the
eigenstates which are directly obtained in our formalism from the
decompositions on the axial basis. In the neutron spectra, our [521]3/2
eigenstate (with a squared overlap of 60 percent) is labeled [512]3/2 by
Bonche \etal{}, and in the proton spectra, their [532]5/2 eigenstate is found
to be in our formalism [303]5/2 (with a squared overlap of 81 percent) while
the [523]5/2 state lies 500 keV below.

Comparing more closely the two results, the level spacings are seen to
slightly differ, leading to different $\Omega$ values for some level
crossings. In some cases, such small effects which we believe to be related to
numerical methods in both cases can appear to be relevant, changing the
properties of a given band. We have checked that these two effects do not
depend on truncation effects by locally performing extended calculations with
$N_0=14$.

\section{\emph{S}-type ellipsoids and superdeformed states in \nuc{150}{Gd}}
\label{sec:styp}

We now wish to present some preliminary results obtained within the
generalized routhian formalism. We use the velocity field of
eq. (\ref{stype-field}) which couples the global rotation of the nucleus with
an intrinsic vortical field and is known after Riemann and Chandrasekhar
\cite{Chandra} as the \emph{S}-type ellipsoid approximation. In this
particular case the expression (\ref{routh-spin}) of the routhian can be
developed through straightforward calculations as the doubly constrained
operator
\begin{equation}
R = H - \Omega (L_1 + s_1) - \omega K_1 ,
\label{routh-stype}
\end{equation}
where $K_1$ is the first component of the Kelvin circulation operator which
writes in cartesian coordinates
\begin{equation}
K_1 = - \cpi\hbar \left( q\, x_2 \frac{\partial}{\partial x_3} - \frac{1}{q}\,
 x_3 \frac{\partial}{\partial x_2} \right) ,
\label{kelv-op}
\end{equation}
$q$ being the axis ratio $a_3/a_2$ defined in Section \ref{ssec:sym}.  This
definition corresponds to a double stretching of the angular momentum operator
in both positions and momenta.

The introduction of the \emph{S}-type velocity field in the context of nuclear
physics has been already discussed by two of the authors and
I.~N.~Mikha\"\i{}lov in Refs.~\cite{Nupha,Staggering}. In particular it has
been stated that the Kelvin circulation defined above, which turns out to be
the conjugate variable associated with the angle whose time derivative is
called $\omega$ , should be quantized as is the angular momentum associated to
the rotation angular velocity $\Omega$. This quantization effect has been
proposed as a tentative explanation of the striking 2~$\hbar$ staggering
phenomenon observed in some superdeformed bands in the $A=150$ mass region
\cite{Haslip} and less clearly in the $A=130$ mass region for Cerium isotopes
\cite{Cerium}.

The expectation value of the Kelvin circulation is easily obtained from the
expression (\ref{obs-angular-last}) of the total angular momentum by
performing the stretching in coordinates and momenta and removing the spin
degrees of freedom contribution, as
\begin{equation}
\langle K_1/\hbar\rangle = \frac{2\pi}{A} \int_0^\infty r\mathrm{d}r
\int_0^\infty \mathrm{d}z \left[ \frac{r}{q} j_z^{(1)} - q z(j_\theta^{(1)} + 
j_r^{(1)}) \right] .
\label{obs-kelv}
\end{equation}
The quantization rule for the Kelvin circulation writes
\begin{equation}
\langle K_1/\hbar \rangle^2 = J(J+1) ,
\label{kelvin-quant}
\end{equation}
where $J$ must be an even integer due to the C2-symmetry character of the
solution.

Upon varying the Kelvin circulation from its yrast value (which corresponds to
a vanishing $\omega$ value), it is possible to obtain a wide class of
collective flows, including irrotational flow, shear modes and others,
corresponding to various moments of inertia differing from the one obtained on
the yrast line. Such effects being also observed whenever pairing
correlations are introduced, we can thus simulate the consequences of pairing
forces in terms of current patterns by shifting the Kelvin circulation from
its yrast value through the second dynamical constraint $-\omega K_1$.

\begin{figure}[t]
\begin{center}
\includegraphics[width=.75\textwidth]{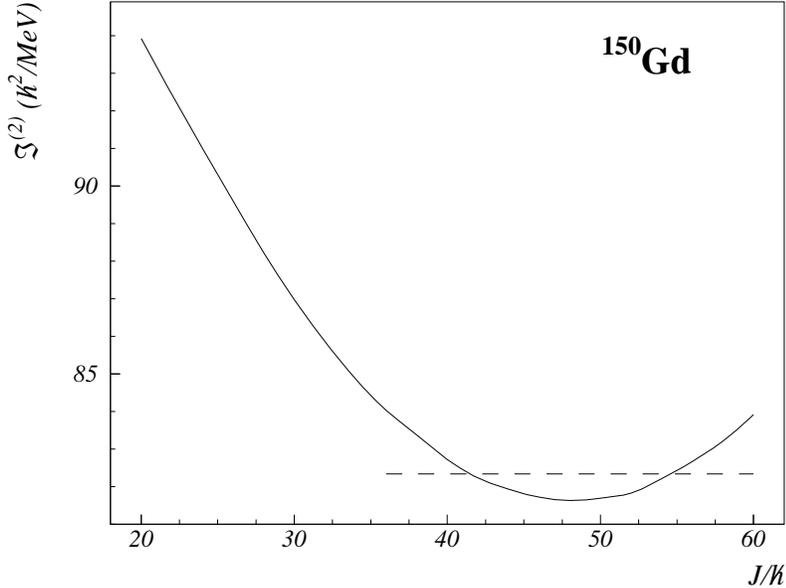}
\label{kelvin}
\caption{Dynamical moment of inertia of $I=50$ $\hbar$ solutions of the nuclei
\nuc{150}{Gd} as a function of the Kelvin circulation $J$. The dashed line
represent the yrast value of the moment of inertia. In all the calculations
reported here we have used the basis parameters optimized for the yrast
$I=50$~$\hbar$ solution.}
\bigskip
\end{center}
\end{figure}

To illustrate this point, we present in Fig.~\ref{kelvin} the variation of the
dynamical moment of inertia as a function of $J$ for \nuc{150}{Gd} solutions
with $I=50$~$\hbar$ (corresponding to $\hbar\Omega=0.544$~MeV on the yrast 
line). The moment of inertia is seen to fit rather nicely on a parabola and
can vary above and below the yrast value as $J$ is varied. More precisely,
classifying the solutions by the rigidity $r$ of their flow patterns defined 
\cite{Rosensteel} as
\begin{equation}
r = 1 + \frac{\omega}{2\Omega} (q+1/q) ,
\label{rigidity}
\end{equation}
where $r=0$ corresponds to purely irrotational flow and $r=1$ corresponds to
global rotation, we can roughly associate the higher moments of inertia with
$r<1$ and the lower ones with $r>1$. It is well known, as shown for instance
by Durand \etal{} in Ref. \cite{Durand} that the introduction of pairing
forces leads to irrotational-like current lines. It is then very promising to
notice that in the case of \nuc{150}{Gd} both pairing correlations (as shown
in Ref. \cite{Bonche-Gado}) and $r<1$ \emph{S}-type solutions correspond to a
higher moment of inertia than the yrast (Hartree--Fock) value.

We will not enter much into details here, leaving for a future publication
\cite{Briefreport} a more complete discussion on this subject. Let us merely
mention that the Kelvin circulation value required to obtain a moment of
inertia close to the experimental value which is around 90 $\hbar^2/$MeV in
the $\hbar\Omega\sim 0.55$ MeV region is $J\sim 25$~$\hbar$. Such a value
corresponds to a ratio $J/I=2$ which is, in the hypothesis of
Ref. \cite{Staggering}, a condition required for a 2~$\hbar$ staggering to
appear.

\section{Conclusions}
\label{sec:conclu}

The aim of this paper was to present a new formalism for solving the
Skyrme--Hartree--Fock problem in the case of a generalized routhian which
breaks both time reversal and axial symmetries. Even though such a kind of
formalism was already existing as we recalled in the introduction, previous
works have only described rotating solutions (\ie{} being restricted to the
single usual cranking case). More importantly, the choice here made to use
axially symmetric basis states, has been shown to reduce the computation time
with respect to the usual choice of triaxial basis states.  The results of our
formalism in the cranking case have been tested against the results available
in the literature in the $A=150$ mass region.  They are in quite satisfactory
agreement for usually considered observables in superdeformed nuclei.

We have also performed preliminary calculations within the \emph{S}-type
ellipsoid approximation. In Ref. \cite{Nupha}, we had tackled the problem in
the limiting case where the mean field is mocked-up by an harmonic
potential. Here we started to lift up this approximation. However, this study
is by no means complete and will be investigated in more details in
forthcoming publications. It already gives us some hints, however, about the
strong connection existing, as predicted, between the pairing correlations and
the intrinsic vortical collective modes.
For this purpose, it is clearly required that our formalism be extended to
include pairing correlations even though some important results can certainly
be obtained already within the Hartree--Fock approximation. This work is
in progress, as we are currently working in upgrading our
formalism to an Hartree--Fock--Bogoliubov approach, using a zero-range pairing
force, treating moreover the complicated problem of projection on good
particle number in the approximate and widely used Lipkin--Nogami scheme (see
\eg{} \cite{Lipkin} for the implementation of this scheme in the framework of
self-consistent calculations).
Such a HFB formalism will also be instrumental in testing the hypothesis of
Ref. \cite{Staggering} on the appearance of a 2~$\hbar$ staggering in some
superdeformed bands. Indeed, since the moment of inertia renormalization
induced by the pairing force is accompanied by a more pronounced irrotational
character of the current patterns, the Kelvin circulation spectra along the
yrast line should be very different in the HF and HFB cases. One should
not therefore consider the $J/I$ value in the pure Hartree--Fock case
as the final word and rather consider it as a good starting point for a more
comprehensive study of high spin nuclear dynamics.

\section*{Acknowledgments}
This work which originates from one of the author's (D. S.) PhD thesis has
benefited from numerous fruitful and illuminating discussions with J. Meyer
which we want to thank here. We are also deply indebted to
I. N. Mikha\"\i{}lov who participated with two of the authors (D. S. and
P. Q.) in the development of the generalized routhian formalism in use in this
paper and of the physics of nuclear intrinsic vortical currents. We are also
grateful to J. Libert and J. Dudek who have been very helpful in allowing us
to compare our computing time with those of other existing approaches.

\appendix

\section{Routhians}
\subsection{Skyrme energy functional and local densities}
\label{app:densities}

The energy functional in the case of time-reversal symmetry breaking writes 
as~\cite{Engel}
\begin{equation}
{\cal H}(\boldsymbol r) = {\cal H}_\mathrm{kin}(\boldsymbol r) + {\cal H}_\mathrm{vol}(\boldsymbol r) + {\cal H}_\mathrm{so}(\boldsymbol r) + 
{\cal H}_\mathrm{odd}(\boldsymbol r) + {\cal H}_\mathrm{coul}(\boldsymbol r) ,
\label{functional}
\end{equation}
where these different terms are given below, omitting their $\boldsymbol r$ 
dependence. The kinetic energy part writes
\begin{equation}
{\cal H}_\mathrm{kin}=\frac{A-1}{A}\frac{\hbar^2}{2m}\tau ,
\label{func-kin}
\end{equation}
taking approximately into account the center of mass energy correction. The
rest is given in terms of the force parameters $B_i$ ($i=1,11$) given in Ref.~\cite{Bonche}
\begin{eqnarray}
{\cal H}_\mathrm{vol} & = & B_1\rho^2 + B_2\sum_q{\rho_q^2} + B_7\rho^{\alpha+2} + B_8\rho^{\alpha}\sum_q{\rho_q^2} \nonumber \\
\label{func-tot}
&& + B_3\rho\tau + B_4\sum_q{\rho_q\tau_q} + B_5\rho\boldsymbol\nabla^2\rho + B_6\sum_q{\rho_q\boldsymbol\nabla^2\rho_q} ,
\end{eqnarray}
\begin{equation}
{\cal H}_\mathrm{so}=B_9\left(\rho\boldsymbol\nabla\cdot\boldsymbol J + \sum_q{\rho_q\boldsymbol\nabla\cdot\boldsymbol J_q}\right) ,
\label{func-so}
\end{equation}
\begin{eqnarray}
{\cal H}_\mathrm{odd} & = & B_9 \left( \boldsymbol\nabla \times \boldsymbol\rho \cdot \boldsymbol j + \sum_q \boldsymbol\nabla \times \boldsymbol\rho_q \cdot \boldsymbol j_q \right) - B_3 \boldsymbol j^2 - B_4 \sum_q \boldsymbol j_q^2 \nonumber \\
\label{func-odd}
&& + B_{12} \rho^\alpha \boldsymbol\rho^2 + B_{13} \rho^\alpha \sum_q \boldsymbol\rho_q^2 + 
B_{10} \boldsymbol\rho^2 + B_{11} \sum_q \boldsymbol\rho_q^2 ,
\end{eqnarray}
in which some small contributions have been neglected as was discussed in 
\cite{Bonche} and
\begin{equation}
{\cal H}_\mathrm{coul} = \frac{1}{2} e^2\rho_p \int \frac{\rho_p({\boldsymbol r'})}
{|\boldsymbol r - {\boldsymbol r'}|} \mathrm{d}{\boldsymbol r'} - \frac{3e^2}{4} \left(\frac{3}{\pi}\right)^{1/3} \rho_p^{4/3}
\label{func-coul}
\end{equation}
using the Slater approximation \cite{Slater} for the exchange term whose
accuracy has been checked in Ref \cite{Titin}. These 
expressions make use of the following local densities
\begin{eqnarray}
\label{ro-dens}
\rho_q & = & \sum_k \Phi_k^* \Phi_k , \\
\label{tau-dens}
\tau_q & = & \sum_k \boldsymbol\nabla\Phi_k^* \cdot \boldsymbol\nabla\Phi_k , \\
\label{dj-dens}
\boldsymbol\nabla \cdot \boldsymbol J_q & = & -\cpi \sum_k \boldsymbol\nabla\Phi_k^* \cdot 
\boldsymbol\nabla \times \boldsymbol\sigma \Phi_k \\
\label{j-dens}
\boldsymbol j_q & = & \frac{1}{2\cpi} \sum_k \Phi_k^* \boldsymbol\nabla\Phi_k - 
\Phi_k \boldsymbol\nabla\Phi_k^* , \\
\label{vro-dens}
\boldsymbol\rho_q & = & \sum_k\Phi_k^* \boldsymbol\sigma \Phi_k ,
\end{eqnarray}
where all the sums run over the occupied states for a given charge state $q$. 
When no charge index is present, we refer to the total density like \eg{}
\begin{equation}
\rho = \sum_q \rho_q
\end{equation}

\subsection{Form factors of the one-body routhian}
\label{app:fields}

The form factors appearing in eq.~(\ref{routh-1body}) are expressed as follows:
\begin{eqnarray}
U_q & = & 2 (B_1\rho + B_2\rho_q) + (B_3\tau + B_4\tau_q) + 2 (B_5\Delta\rho + 
B_6\Delta\rho_q) \nonumber\\
&& + (2 + \alpha )B_7 \rho^{1+\alpha} + B_8\rho^{\alpha-1} \left( 2\rho\rho_q 
+ \alpha\sum_q\rho_q^2 \right) + B_9 (\boldsymbol\nabla \cdot \boldsymbol J + \boldsymbol\nabla 
\cdot \boldsymbol J_q) \nonumber\\
\label{scalar-pot}
&& + \alpha\rho^{\alpha-1} \left (B_{12}\boldsymbol \rho^2 + B_{13}\sum_q\boldsymbol \rho_q^2 
\right) + \delta_{qp} \left( V_\mathrm{coul}^\mathrm{dir} - e^2 \left[ \frac{3}{\pi}\rho_p 
\right]^{1/3} \right) ,
\end{eqnarray}
\begin{equation}
f_q = 1 + \frac{2m}{\hbar} (B_3\rho + B_4\rho_q) ,
\label{effect-mass}
\end{equation}
\begin{equation}
\hbar\boldsymbol\alpha_q = \hbar\boldsymbol\beta + 2 (B_3\boldsymbol j + B_4\boldsymbol j_q) - B_9 
(\boldsymbol\nabla \times \boldsymbol\rho + \boldsymbol\nabla \times \boldsymbol\rho_q) ,
\label{alpha-field}
\end{equation}
\begin{eqnarray}
\hbar\boldsymbol S_q & = & \frac{\hbar}{2}\boldsymbol\Omega - B_9 (\boldsymbol\nabla \times \boldsymbol j + 
\boldsymbol\nabla \times \boldsymbol j_q) - 2 (B_{10}\boldsymbol\rho + B_{11}\boldsymbol\rho_q) \nonumber\\
\label{spin-field}
&& - 2\rho^\alpha (B_{12}\boldsymbol\rho + B_{13}\boldsymbol\rho_q) ,
\end{eqnarray}
\begin{equation}
V_q^{\mathrm{so}} = -\frac{B_9}{\hbar^2} (\rho + \rho_q) ,
\label{so-pot}
\end{equation}
and
\begin{equation}
V_{\mathrm{coul}}^\mathrm{dir} = -e^2 \int \frac{\rho_p (\boldsymbol{r'})}{|\boldsymbol r -
\boldsymbol{r'}|} \mathrm{d}\boldsymbol{r'} .
\label{coul-pot}
\end{equation}

\section{Expression in the space coordinates of densities and energy form 
factors}
\subsection{Densities calculation}
\label{app:denscalc}

The calculation of the densities and routhian matrix elements requires 
the knowledge of the derivatives of those functions with respect to the 
three cylindrical coordinates. The derivation with respect to $\theta$ is 
straightforward from equation (\ref{spineur}), whereas the other 
derivatives are given with obvious notations by
\begin{eqnarray}
\label{nabla-r}
\partial_r \Psi^\pm (\boldsymbol r) = \left[ 
\frac{\beta_z\beta_\perp^4}{2\pi} \textrm{e}^{-(\xi^2+\eta)} \right]^{1/2} 
&& \sum_\mu(\pm)^{n_z} C_\mu^k \expe^{\pm i\Lambda\theta}  
\eta^{(|\Lambda|-1)/2} H_{n_z} \bar L_{n_r}^{|\Lambda|} , \\
\label{nabla-z}
\partial_z \Psi^\pm (\boldsymbol r) = \left[ 
\frac{\beta_z^3\beta_\perp^2}{2\pi} \textrm{e}^{-(\xi^2+\eta)} \right]^{1/2} 
&& \sum_\mu(\pm)^{n_z} C_\mu^k \expe^{\pm i\Lambda\theta} \eta^{|\Lambda|} 
\bar H_{n_z} L_{n_r}^{|\Lambda|} ,
\end{eqnarray}
using as was done in Ref.~\cite{Vautherin} the definitions
\begin{equation}
\bar H_{n_z}(\xi) = \xi H_{n_z}(\xi) - H_{n_z+1}(\xi) ,
\end{equation}
\begin{equation}
\bar L_{n_r}^{|\Lambda|} (\eta) = 2 (n_r + 1) L_{n_r+1}^{|\Lambda|} (\eta) -
(2n_r + |\Lambda| + 2 - \eta) L_{n_r} (\eta) .
\end{equation}

From the expressions given by Vautherin (eqs.~(5.5) of Ref.~\cite{Vautherin}) for the $\tau$ and 
$\boldsymbol\nabla^2\rho$ densities in the case of time-reversal symmetry 
and the calculation of $\rho$ performed in sect.~\ref{ssec:sym} we can easily 
obtain in our case the following expressions:
\begin{eqnarray}
\tau & = & \frac{\beta_z\beta_\perp^2}{\pi} \textrm{e}^{-(\xi^2+\eta)}
\sum_{\mu,\mu'} \rho_{\mu\mu'}
\eta^{(|\Lambda|+|\Lambda'|)/2-1} \delta^{2p}_{\Lambda-\Lambda'} 
\cos [(\Lambda-\Lambda')\theta] \nonumber \\
\label{tau-dens-2}
&& \left\{ \eta\beta_z^2 \bar H_{n_z} \bar H_{n'_z} L_{n_r}^{|\Lambda|}
L_{n'_r}^{|\Lambda'|} + \beta_\perp^2 H_{n_z} H_{n'_z} \left[ 
\bar L_{n_r}^{|\Lambda|} \bar L_{n'_r}^{|\Lambda'|} + \Lambda\Lambda' 
L_{n_r}^{|\Lambda|} L_{n'_r}^{|\Lambda'|} \right] \right\} ,
\end{eqnarray}
\begin{eqnarray}
\boldsymbol\nabla^2\rho & = & 2\tau_q \frac{2\beta_z\beta_\perp^2}{\pi} \textrm{e}^{-(\xi^2+\eta)}
\sum_{\mu,\mu'} \rho_{\mu\mu'}
\eta^{(|\Lambda|+|\Lambda'|)/2} \delta^{2p}_{\Lambda-\Lambda'} 
\cos [(\Lambda-\Lambda')\theta] \nonumber \\
&& \left\{ \beta_z^2 \left[\xi^2-2(n_z+\half)\right] + \beta_\perp^2 \left[\eta-
2(2n_r+|\Lambda|+1)\right] \right\}\nonumber \\
\label{dro-dens}
&& H_{n_z} H_{n'_z} L_{n_r}^{|\Lambda|} L_{n'_r}^{|\Lambda'|} .
\end{eqnarray}
Using the expression (3.7) of Ref.~\cite{Vautherin} for the cylindrical 
components of the operator $\boldsymbol\nabla \times \boldsymbol\sigma$, we obtain here 
the divergence of the spin-orbit density as
\begin{eqnarray}
\boldsymbol\nabla\cdot\boldsymbol J & = & 2 \sum_k \Im \left\{ \partial_r{\Psi_k^+}^* 
\frac{1}{r}\partial_\theta\Psi_k^+ - \partial_r{\Psi_k^-}^* 
\frac{1}{r}\partial_\theta\Psi_k^- \right\} \nonumber \\
&& + \Im \left\{ s \left( e^{-\cpi\theta}\frac{1}{r}\partial_\theta{\Psi_k^+}^* 
\partial_z\Psi_k^- + e^{\cpi\theta}\frac{1}{r}\partial_\theta{\Psi_k^-}^* 
\partial_z\Psi_k^+ \right) \right\}\nonumber \\
\label{dj-dens-2}
&& + \Re \left\{ s \left(e^{-\cpi\theta}\partial_r{\Psi_k^+}^*\partial_z\Psi_k^-
- e^{\cpi\theta}\partial_r{\Psi_k^-}^*\partial_z\Psi_k^+ \right) \right\} ,
\end{eqnarray}
where $\Re\{x\}$ ($\Im\{x\}$ resp.) represents the real part (imaginary part resp.) 
of $x$. This expression may be worked out with the help of the equations 
(\ref{spineur}), (\ref{nabla-r}) and (\ref{nabla-z}) as
\begin{eqnarray}
\boldsymbol\nabla \cdot &\boldsymbol J & = \frac{\beta_z\beta_\perp^2}{\pi}\expe^{-
(\xi^2+\eta)} \sum_{\mu,\mu'} \rho_{\mu\mu'} \eta^{(|\Lambda|+|\Lambda'|)/2}
\nonumber \\
\label{dj-dens-3}
&& \beta_\perp^2\eta^{-1}\Lambda' H_{n_z} H_{n'_z} \bar L_{n_r}^{|\Lambda|} 
L_{n'_r}^{|\Lambda'|} \cos [(\Lambda-\Lambda')\theta] \left(1 + 
(-)^{n_z+n'_z}\right) \\
&& - s \beta_z\beta_\perp\eta^{-1/2} H_{n_z} \bar H_{n'_z} \Lambda 
L_{n_r}^{|\Lambda|} L_{n'_r}^{|\Lambda'|} \cos [(\Lambda+\Lambda'+1)\theta]
\left( (-)^{n'_z} - (-)^{n_z} \right) \nonumber \\
&& + s \beta_z\beta_\perp\eta^{-1/2} H_{n_z} \bar H_{n'_z} \bar 
L_{n_r}^{|\Lambda|} L_{n'_r}^{|\Lambda'|} \cos [(\Lambda+\Lambda'+1)\theta] 
\left( (-)^{n'_z} - (-)^{n_z} \right) . \nonumber 
\end{eqnarray}
Now introducing the definition
\begin{equation}
\delta^{2p+1}_{\Lambda-\Lambda'} = \begin{cases}
 1 &\quad \text{if }\Lambda-\Lambda' \text{ is an odd integer} , \\
0 &\quad \text{if }\Lambda-\Lambda' \text{ is an even integer} ,
\end{cases}
\label{cronecker-2}
\end{equation}
we finally write this density in the following way:
\begin{eqnarray}
\boldsymbol\nabla \cdot \boldsymbol J & = & \frac{2\beta_z\beta_\perp^3}{\pi}\expe^{-
(\xi^2+\eta)} \sum_{\mu,\mu'} \rho_{\mu\mu'} \eta^{(|\Lambda|+|\Lambda'|-1)/2}
\nonumber \\
\label{dj-dens-last}
&& \Lambda'\beta_\perp\eta^{-1/2} H_{n_z} H_{n'_z} \bar L_{n_r}^{|\Lambda|} 
L_{n'_r}^{|\Lambda'|} \delta^{2p}_{\Lambda-\Lambda'} \cos [(\Lambda-\Lambda')\theta]
\\
&& + s(-)^{n'z} \beta_z H_{n_z} \bar H_{n'_z} L_{n'_r}^{|\Lambda'|} \left( 
\bar L_{n_r}^{|\Lambda|} - \Lambda L_{n_r}^{|\Lambda|} \right) 
\delta^{2p+1}_{\Lambda-\Lambda'} \cos [(\Lambda+\Lambda'+1)\theta] . \nonumber 
\end{eqnarray}

From the definition (\ref{j-dens}) of the current density, one can write the 
$r$ component of this density like
\begin{equation}
j_r = \sum_k \Im \left\{ {\Psi_k^+}^*\partial_r\Psi_k^+ + 
{\Psi_k^-}^*\partial_r\Psi_k^- \right\} ,
\label{jr-dens}
\end{equation}
which can be readily developed as
\begin{eqnarray}
j_r & = & \frac{\beta_z\beta_\perp^2}{2\pi}\expe^{-(\xi^2+\eta)} 
\sum_{\mu,\mu'} \rho_{\mu\mu'} \eta^{(|\Lambda|+|\Lambda'|)/2} 
\beta_\perp \eta^{-1/2} H_{n_z} H_{n'_z} L_{n_r}^{|\Lambda|} 
\bar L_{n'_r}^{|\Lambda'|} \nonumber \\
\label{jr-dens-2}
&& \sin [(\Lambda'-\Lambda)\theta] \left(1 - (-)^{n_z+n'_z}\right) ,
\end{eqnarray}
since the sine is an odd function. This expression can be simplified as
\begin{eqnarray}
j_r & = & - \frac{\beta_z\beta_\perp^3}{\pi}\expe^{-(\xi^2+\eta)} 
\sum_{\mu,\mu'} \rho_{\mu\mu'} \eta^{(|\Lambda|+|\Lambda'|-1)/2} 
\delta^{2p+1}_{\Lambda-\Lambda'} \sin [(\Lambda-\Lambda')\theta] \nonumber \\
\label{jr-dens-last}
&& H_{n_z} H_{n'_z} L_{n_r}^{|\Lambda|} \bar L_{n'_r}^{|\Lambda'|} .
\end{eqnarray}
In the same way, the two other components of this density are expressed as
\begin{eqnarray}
j_\theta & = & \frac{\beta_z\beta_\perp^3}{\pi}\expe^{-(\xi^2+\eta)} 
\sum_{\mu,\mu'} \rho_{\mu\mu'} \eta^{(|\Lambda|+|\Lambda'|-1)/2} 
\delta^{2p+1}_{\Lambda-\Lambda'} \cos [(\Lambda-\Lambda')\theta] \nonumber \\
\label{jt-dens-last}
&& \Lambda' H_{n_z} H_{n'_z} L_{n_r}^{|\Lambda|} \bar L_{n'_r}^{|\Lambda'|} ,
\end{eqnarray}
\begin{eqnarray}
j_z & = & - \frac{\beta_z^2\beta_\perp^2}{\pi}\expe^{-(\xi^2+\eta)} 
\sum_{\mu,\mu'} \rho_{\mu\mu'} \eta^{(|\Lambda|+|\Lambda'|-1)/2} 
\delta^{2p+1}_{\Lambda-\Lambda'} \sin [(\Lambda-\Lambda')\theta] \nonumber \\
\label{jz-dens-last}
&& H_{n_z} \bar H_{n'_z} L_{n_r}^{|\Lambda|} L_{n'_r}^{|\Lambda'|} .
\end{eqnarray}
To calculate the spin-vector density, we will use the cylindrical coordinates 
of the operator $\boldsymbol\sigma$ which are given in terms of the Pauli matrices 
by
\begin{equation}
\begin{array}{rcl}
\sigma_r & = & \frac{1}{2} \left( \sigma_+ \expe^{-\cpi\theta} + \sigma_- 
\expe^{\cpi\theta} \right) , \\
\sigma_\theta & = & - \frac{\cpi}{2} \left( \sigma_+ \expe^{-\cpi\theta} - 
\sigma_- \expe^{\cpi\theta} \right) , 
\end{array}
\label{pauli}
\end{equation}
the $z$ component being evident. We can then write
\begin{eqnarray}
\label{ror-dens}
\rho_r & = & 2 \sum_k \Re \left\{ s {\Psi_k^+}^*\Psi_k^- \expe^{-\cpi\theta}
\right\} , \\
\label{rot-dens}
\rho_\theta & = & 2 \sum_k \Im \left\{ s {\Psi_k^+}^*\Psi_k^- 
\expe^{-\cpi\theta} \right\} , \\
\label{roz-dens}
\rho_z & = &  \sum_k {\Psi_k^+}^*\Psi_k^+ - {\Psi_k^-}^*\Psi_k^- .
\end{eqnarray}
It is then straightforward to obtain
\begin{eqnarray}
\rho_r & = & \frac{\beta_z\beta_\perp^2}{\pi}\expe^{-(\xi^2+\eta)} 
\sum_{\mu,\mu'} s (-)^{n'_z} \rho_{\mu\mu'} \eta^{(|\Lambda|+|\Lambda'|)/2} 
\delta^{2p}_{\Lambda-\Lambda'} \cos [(\Lambda+\Lambda'+1)\theta] \nonumber \\
\label{ror-dens-2}
&& H_{n_z} H_{n'_z} L_{n_r}^{|\Lambda|} L_{n'_r}^{|\Lambda'|} ,
\end{eqnarray}
since the sum of the $\mu\mu'$ and $\mu'\mu$ terms cancel when $n_z+n'_z$ 
is an odd integer. The $\theta$ component is simply given by
\begin{eqnarray}
\rho_\theta & = & - \frac{\beta_z\beta_\perp^2}{\pi}\expe^{-(\xi^2+\eta)} 
\sum_{\mu,\mu'} s (-)^{n'_z} \rho_{\mu\mu'} \eta^{(|\Lambda|+|\Lambda'|)/2} 
\delta^{2p}_{\Lambda-\Lambda'} \sin [(\Lambda+\Lambda'+1)\theta] \nonumber \\
\label{rot-dens-2}
&& H_{n_z} H_{n'_z} L_{n_r}^{|\Lambda|} L_{n'_r}^{|\Lambda'|} ,
\end{eqnarray}
and since equations (\ref{roz-dens}) and (\ref{ro-dens-2}) only differ by 
a minus sign between the two terms of the right-hand side we have
\begin{eqnarray}
\rho_z & = & \frac{\beta_z\beta_\perp^2}{\pi} \textrm{e}^{-(\xi^2+\eta)}
\sum_{\mu,\mu'} \rho_{\mu\mu'}
\eta^{(|\Lambda|+|\Lambda'|)/2} \delta^{2p+1}_{\Lambda-\Lambda'} 
\cos [(\Lambda-\Lambda')\theta] \nonumber \\
\label{roz-dens-2}
&& H_{n_z} H_{n'_z} L_{n_r}^{|\Lambda|} L_{n'_r}^{|\Lambda'|} . 
\end{eqnarray}
The components of the curl of $\boldsymbol j$ and $\boldsymbol\rho$ which also 
enter the one-body routhian can be obtained upon mixing partial 
derivatives in the following way:
\begin{eqnarray}
\curl_r \boldsymbol j & = & \partial_\theta j_z - \partial_z j_\theta , \\
\curl_\theta \boldsymbol j & = & \partial_z j_r - \partial_r j_z , \\
\curl_z \boldsymbol j & = & \partial_r j_\theta - \partial_\theta j_r
+ \frac{1}{r} j_\theta .
\end{eqnarray}

\subsection{Calculation of the Coulomb potential}
\label{app:coul}

It is well known and used in Ref. \cite{Vautherin} that the Coulomb potential
of equation (\ref{coul-pot}) can be written after two integrations by part as
\begin{equation}
V_\mathrm{coul}^\mathrm{dir}(\boldsymbol r) = - \frac{e^2}{2} \int |\boldsymbol r - 
\boldsymbol{r'}| \boldsymbol\nabla^2 \rho_p (\boldsymbol{r'}) \;\mathrm{d}\boldsymbol{r'} .
\label{cou-pot-2}
\end{equation}
Developing
\begin{equation}
|\boldsymbol r - \boldsymbol{r'}| = \left[ (z-z')^2 + (r+r')^2 \right]^{1/2} 
\left[1 - k^2\cos^2 \left(\frac{\theta-\theta'}{2}\right) \right]^{1/2} ,
\end{equation}
with
\begin{equation}
k^2 = \frac{4rr'}{(z-z')^2+(r+r')^2} ,
\end{equation}
and writing as suggested by Table~\ref{dens-decomp}
\begin{equation}
\boldsymbol\nabla^2\rho_p (\boldsymbol{r'}) = \sum_{n \geq 0} D^{2n}(r',z') 
\cos 2n\theta' ,
\end{equation}
the Coulomb potential can be developed in the following way
\begin{eqnarray}
V_\mathrm{coul}^\mathrm{dir} (\boldsymbol r) & = & \frac{e^2}{2} \sum_{n \geq 0}
\int_0^\infty r'\mathrm{d}r' \int_{-\infty}^\infty \mathrm{d}z' \left[
(z-z')^2 + (r+r')^2 \right]^{1/2} D^{2n}(r',z') \nonumber \\
\label{coul-pot-3}
&& \int_0^{2\pi} \mathrm{d}\theta' \left[1 - k^2\cos^2 
\left(\frac{\theta-\theta'}{2}\right) \right]^{1/2} \cos 2n\theta' .
\end{eqnarray}
In terms of $u=(\theta-\theta')/2$, the angular part of the above 
integral rewrites as
\begin{equation}
I^{2n} = 2 \int_0^\pi \mathrm{d}u \left(1 - k^2\cos^2u\right)^{1/2}
\cos (4nu - 2n\theta).
\end{equation}
Developing the last cosine of this expression, we can factorize the 
$\theta$ dependence out of the integral and obtain
\begin{equation}
I^{2n} = 4 \cos 2n\theta \int_0^{\pi/2} \mathrm{d}u \left(1 - 
k^2\cos^2u\right)^{1/2} \cos 4nu ,
\end{equation}
since the other term is vanishing. We can then write the Coulomb potential 
as a Fourier series in the $\theta$ variable:
\begin{eqnarray}
V_\mathrm{coul}^\mathrm{dir} (\boldsymbol r) & = & 4e^2 \sum_{n \geq 0} \cos 2n
\theta \int_0^\infty r'\mathrm{d}r' \int_0^\infty \mathrm{d}z' \left[ (z-z')^2
+ (r+r')^2 \right]^{1/2} \nonumber \\
\label{coul-pot-last}
&& D^{2n}(r',z') \int_0^{\pi/2} \mathrm{d}u \left(1 - k^2\cos^2 u \right)^{1/2} 
\cos 4nu ,
\end{eqnarray}
where we have restricted the integral over $z$ to positive values using
parity properties. The $r'$ and $z'$ integrals are then performed numerically
through 10 points Gauss--Hermite and Gauss--Laguerre quadrature formulas
whereas the $u$ integral is performed through a 48 points Gauss--Legendre
quadrature formula during the first iteration, the results being then stored for 
future use.

\section{Calculations of routhian matrix elements}
\label{app:matrix}

Comparing our expression (\ref{mat-U-last}) for the scalar part of the
routhian matrix element with Vautherin's equation (4.22) in
Ref.~\cite{Vautherin}, we can straightforwardly obtain the kinetic part of the
matrix element as
\begin{eqnarray}
\langle\mu s |  \boldsymbol\nabla f \cdot \boldsymbol\nabla | \mu' s 
\rangle &=& - (1 + \delta_{\Lambda-\Lambda'}^0) \int_0^\infty 
\mathrm{d}\eta \int_0^{\infty} \mathrm{d}\xi \expe^{-(\xi^2+\eta)} 
\eta^{(|\Lambda|+|\Lambda'|)/2} \nonumber \\
&& f^{(|\Lambda-\Lambda'|)} \left[ \beta_\perp^2 \eta^{-1} 
H_{n_z} H_{n'_z} \left(\bar L_{n_r}^{|\Lambda|} \bar L_{n'r}^{|\Lambda'|} + 
\Lambda\Lambda' L_{n_r}^{|\Lambda|} L_{n'r}^{|\Lambda'|} \right) \right. 
\nonumber \\
\label{mat-f-last}
&& \left. + \beta_z^2 L_{n_r}^{|\Lambda|} L_{n'r}^{|\Lambda'|} \bar H_{n_z} 
\bar H_{n'_z} \right] ,
\end{eqnarray}
where we use, as will be done throughout this Appendix, the notation 
$\Gamma^{(n)}$ to represent the $n$-th non-vanishing component of the Fourier 
series of any routhian form factor $\Gamma$ (see Table~\ref{pot-decomp}).
The calculation of the $\boldsymbol\alpha$ part of the matrix element can be
rewritten through an integration by parts as 
\begin{equation}
\langle \mu s | (\boldsymbol\alpha \cdot \boldsymbol\nabla + \boldsymbol\nabla \cdot \boldsymbol\alpha) |
\mu' s \rangle = \langle \mu s | \boldsymbol\alpha \cdot \boldsymbol\nabla | \mu' s \rangle -
\langle \mu' s | \boldsymbol\alpha \cdot \boldsymbol\nabla | \mu s \rangle^* ,
\label{alpha-by-part}
\end{equation}
which allows us to only calculate the first term on the right-hand side, the 
second term being easily deduced from it. 
\begin{eqnarray}
\langle \mu s | \boldsymbol\alpha \cdot \boldsymbol\nabla | \mu' s \rangle & = & \frac{1}{2}
\sum_{n\geq 0} \Big( \langle \mu | \cos [(2n+1)\theta]\,
\alpha_\theta^{(2n+1)} \frac{\partial_\theta}{r} | \mu' 
\rangle \nonumber \\
&& \quad + \langle \mu | \sin [(2n+1)\theta]\,
\big(\alpha_r^{(2n+1)} \partial_r + \alpha_z^{(2n+1)} \partial_z\big) | \mu' 
\rangle \Big) \nonumber\\
\label{alpha-1}
&& + (-)^{n_z+n'_z} \big[ (\mu,\mu') \rightarrow (\bar\mu,\bar\mu') \big] ,
\end{eqnarray}
where the term $\big[(\mu,\mu') \rightarrow (\bar\mu,\bar\mu')\big]$ represents
a term identical to the previous one with $\bar\mu$ and $\bar\mu'$ replacing
$\mu$ and $\mu'$. Making use of equation (\ref{select-rule}) and of the
following identity:
\begin{equation}
\int_0^{2\pi} \expe^{\cpi k\theta} \sin m\theta = \begin{cases}
0 \phantom{\cpi \pi \frac{k}{m}} &\quad \text{if } m \neq
|k| \text{ and } m\neq0 ,  \\
\cpi \pi \frac{k}{m} \phantom{0} &\quad \text{if } m = |k| ,
\end{cases}
\label{select-rule-sin}
\end{equation}
one can notice that the $\mu\mu'$ and $\bar\mu\bar\mu'$ contributions in 
equation (\ref{alpha-1}) are identical. The sum over $n$ therefore reduces to
\begin{eqnarray}
\langle \mu s | \boldsymbol\alpha \cdot \boldsymbol\nabla | \mu' s \rangle & = & \langle \mu
| \sin (|\Lambda-\Lambda'|\theta)
\big(\alpha_r^{(|\Lambda-\Lambda'|)} \partial_r + \alpha_z^{(|\Lambda-\Lambda'|)}
\partial_z\big) | \mu' \rangle \nonumber\\
\label{alpha-2}
&& + \langle \mu | \cos [|\Lambda-\Lambda'|\theta]\,
\alpha_\theta^{(|\Lambda-\Lambda'|)} \frac{\partial_\theta}{r} | 
\mu' \rangle ,
\end{eqnarray}
which expands as
\begin{eqnarray}
\langle \mu s | \boldsymbol\alpha \cdot \boldsymbol\nabla | \mu' s \rangle & = &
\cpi \beta_\perp \int_0^\infty \mathrm{d}\eta \int_0^\infty
\mathrm{d}\xi \expe^{-(\xi^2+\eta)} \eta^{(|\Lambda| + |\Lambda'|-1)/2}
H_{n_z} L_{n_r}^{|\Lambda|} \nonumber \\
&&\bigg[ \frac{\Lambda'-\Lambda}{|\Lambda'-\Lambda|} \left(
\alpha_r^{(|\Lambda-\Lambda'|)} H_{n'_z} \bar L_{n'_r}^{|\Lambda'|} +
\frac{\beta_z}{\beta_\perp} \eta^{1/2} \alpha_z^{(|\Lambda-\Lambda'|)} \bar
H_{n'_z} L_{n'_r}^{|\Lambda'|} \right) \nonumber \\
\label{alpha-last}
&& + \alpha_\theta^{(|\Lambda-\Lambda'|)}\Lambda' H_{n'_z}
L_{n'_r}^{|\Lambda'|} \bigg]
\end{eqnarray}
The spin part of the matrix element involves a spin diagonal term and two
spin anti-diagonal terms. Using the relations (\ref{pauli}) and performing the
spin part of the scalar product, they respectively write
\begin{eqnarray}
\langle \mu s | S_z \sigma_z | \mu' s \rangle & = & \frac{1}{2} \sum_{n\geq 0} 
\langle \mu | \cos [(2n+1)\theta]\, S_z^{(2n+1)} | \mu' \rangle \nonumber \\
\label{spin-z-1}
&& - (-)^{n_z+n'_z} \langle \bar\mu | \cos [(2n+1)\theta]\, S_z^{(2n+1)} | 
\bar\mu' \rangle ,
\end{eqnarray}
\begin{eqnarray}
\langle \mu s | S_r \sigma_r  | \mu' s \rangle & = & \frac{1}{2}
\sum_{n\geq 0} s(-)^{n'_z}\langle\mu| \expe^{-\cpi\theta} \cos [(2n+1)\theta]\,
S_r^{(2n+1)} |\bar\mu'\rangle \nonumber \\
&& + s(-)^{n_z} \langle\bar\mu| \expe^{\cpi\theta} \cos [(2n+1)\theta]\,
S_r^{(2n+1)} |\mu'\rangle ,
\end{eqnarray}
and
\begin{eqnarray}
\langle \mu s | S_\theta \sigma_\theta | \mu' s \rangle & = & \frac{-\cpi}{2}
\sum_{n\geq 0} s(-)^{n'_z}\langle\mu| \expe^{-\cpi\theta} \sin [(2n+1)\theta]\, 
S_\theta^{(2n+1)} |\bar\mu'\rangle \nonumber \\
&& - s(-)^{n_z}\langle\bar\mu| \expe^{\cpi\theta} [\sin(2n+1)\theta]\, 
S_\theta^{(2n+1)} |\mu'\rangle.
\end{eqnarray}
In all these three expressions again, the two terms on the right-hand side 
are the same and we finally obtain the expression
\begin{eqnarray}
\langle \mu s | \boldsymbol S \cdot \boldsymbol\sigma | \mu' s \rangle & = & 
\int_0^\infty \mathrm{d}\eta \int_0^\infty \mathrm{d}\xi 
\expe^{-(\xi^2+\eta)} \eta^{(|\Lambda| + |\Lambda'|)/2} H_{n_z} H_{n'_z}
L_{n_r}^{|\Lambda|} L_{n'_r}^{|\Lambda'|} \nonumber \\
&& \Big[ S_z^{(|\Lambda-\Lambda'|)}
+ s (-)^{n_z} \big( S_r^{(|\Lambda+\Lambda'+1|)} \nonumber \\
&& - \frac{\Lambda+\Lambda'+1}
{|\Lambda+\Lambda'+1|} S_\theta^{(|\Lambda+\Lambda'+1|)} \big) \Big].
\end{eqnarray}
One can easily develop the spin-orbit term of the one-body routhian of eq. (\ref{routh-1body}) as
\begin{eqnarray}
U^{\mathrm{so}} = ( \boldsymbol\nabla V^{\mathrm{so}} \times \boldsymbol\nabla ) \cdot \boldsymbol\sigma & = & (\partial_r V^{\mathrm{so}}) \big(\frac{\partial_\theta}{r} \sigma_z -
\partial_z \sigma_\theta\big) + \frac{1}{r} (\partial_\theta V^{\mathrm{so}})
\big(\partial_z\sigma_r - \partial_r\sigma_z\big) \nonumber \\
\label{sporb-1}
&& + (\partial_z V^{\mathrm{so}}) \big(\partial_r\sigma_\theta - 
\frac{\partial_\theta}{r}\sigma_r\big) .
\end{eqnarray}
As the calculation of the matrix element is rather long, we will not enter into
details here. It has been established in Ref.~\cite{MaThese} using the same
technique as we used in this Appendix that, after integrating by parts
to eliminate the $(\partial_i V^{\mathrm{so}})$ terms, we get:
\begin{eqnarray}
\langle \mu s | U^{\mathrm{so}} | \mu' s \rangle & = &
\beta_\perp \int_0^\infty \mathrm{d}\eta \int_0^\infty
\mathrm{d}\xi \expe^{-(\xi^2+\eta)} \eta^{(|\Lambda| + |\Lambda'|)/2-1}
\bigg\{ \nonumber \\
&& \beta_\perp (1 + \delta_{|\Lambda-\Lambda'|}^0)
V_{\mathrm{so}}^{(|\Lambda-\Lambda'|)} H_{n_z} H_{n'_z} \big[ \Lambda
L_{n_r}^{|\Lambda|} \bar L_{n'_r}^{|\Lambda'|} + \Lambda' \bar
L_{n_r}^{|\Lambda|} L_{n'_r}^{|\Lambda'|}\big] \nonumber \\
&& -s (-)^{n_z} \beta_z \eta^{1/2} (1 + 
\delta_{|\Lambda+\Lambda'+1|}) V_{\mathrm{so}}^{(|\Lambda+\Lambda'+1|)}\big[
H_{n_z} \bar H_{n'_z} L_{n'_r}^{|\Lambda'|} \nonumber \\
\label{sporb-last}
&& (\bar L_{n_r}^{|\Lambda|} - \Lambda
L_{n_r}^{|\Lambda|}) - \bar H_{n_z} H_{n'_z} L_{n_r}^{|\Lambda|} (\bar L_{n'_r}^{|\Lambda'|} - \Lambda'
L_{n'_r}^{|\Lambda'|}) \big] \bigg\}
\end{eqnarray}

\end{document}